\documentclass[12pt]{article}
\usepackage{amsmath}
\usepackage{amssymb}
\usepackage{graphicx}
\usepackage{float}
\usepackage{indentfirst}
\usepackage{mathrsfs}
\usepackage{dsfont}

\usepackage{setspace}

\usepackage[labelfont=bf,labelsep=period]{caption}

\usepackage[numbers,sort&compress]{natbib}

\newtheorem{proposition}{Proposition}
\newtheorem{corollary}{Corollary}
\newtheorem{lemma}{Lemma}

\newtheorem{conjecture}{Conjecture}

\title{The Diminished  Quantum Uncertainty in Multipartite Entanglement\\ [0.7cm]}

\author
{Jun-Li Li and Cong-Feng Qiao$^{\ast}$ \\ [0.2cm]
\normalsize{Department of Physics, University of Chinese Academy of Sciences,} \\
\normalsize{YuQuan Road 19A, Beijing 100049, China}\\[2pt]
\normalsize{Key Laboratory of Vacuum Physics, University of Chinese Academy of Sciences}\\
\normalsize{YuQuan Road 19A, Beijing 100049, China} \\ [3mm]
\normalsize{$^\ast$ To whom correspondence should be addressed; E-mail: qiaocf@ucas.ac.cn.}
}

\date{}

\begin{document}
\baselineskip24pt \maketitle
\begin{abstract}\doublespacing
The uncertainty principle and entanglement are two fundamental, but yet not well understood, features of quantum theory. The uncertainty relation reflects the capability limit in acquiring the knowledge of different physical properties of a particle simultaneously, while on the other side, the quantum entanglement renders the entangled quanta lose their independence, including measurements imposed on them. By virtue of the majorization, here we establish a general correlation relation for quantum uncertainty and multipartite entanglement. Within this scheme, the optimization problems for entropy and majorization uncertainty relation are solved. We obtain a diminished uncertainty relation in the presence of multipartite entanglement, where the lower bound is connected with the entanglement class. This result is inspiring, reveals the intrinsic quantitative connection between uncertainty relation and entanglement, and may have a deep impact on quantum measurement in application.
\end{abstract}

\newpage

\section{Introduction}

The uncertainty principle is one of the few extraordinary features distinguishing quantum theory from classical ones. The idea of indeterminacy was first proposed by Heisenberg in form of $\Delta p \Delta q\sim h$, where $h$ is the Planck constant and $\Delta p$ and $\Delta q$ represent the precisions in determining the canonical conjugate observables $p$ and $q$ \cite{Heisenberg-o}. In the literature, whereas the most representative uncertainty relation is the Heisenberg-Robertson one \cite{Robertson}:
\begin{equation}
\Delta A^2 \Delta B^2 \geq \frac{1}{4}|\langle [A,B] \rangle|^2 \; , \label{Robertson-Uncertainty}
\end{equation}
where $\Delta X^2 = \langle X^2\rangle - \langle X\rangle^2$ is the variance and commutator defined as $[A,B] \equiv AB-BA$. Equation (\ref{Robertson-Uncertainty}) may apply to arbitrary pair of incompatible observables in quantum systems. Unsatisfied with this measurement incapability about incompatible observables, Einstein and his collaborators questioned the completeness of quantum mechanics by constructing an ad hoc state \cite{EPR}, the entangled state. Since then ongoing efforts have been devoted to the study of entanglement and its nonlocal phenomena, and eventually lead to the development of quantum information sciences, which are now responsible for the so called ``second quantum revolution" \cite{Second-Rev}.

Nowadays, the prevailing uncertainty relation exhibits in two different kinds of forms, the variance and entropy based ones, which are shown recently to be interconvertible \cite{Equiv-V-E}. A lasting criticism on variance-based uncertainty relation is about its lower bound state dependence \cite{Entropy-0}. Though it is possible to get a state independent lower bound with the help of Bloch vectors \cite{Reformulating-Li, State-independent-Qian}, the state independent uncertainty relations involve generically complex variance functions \cite{TUR}. The entropic uncertainty relation is designated having state independent lower bounds, but the optimization of the lower bounds are difficult for general observable and high dimensional system, i.e. certain numerical method is necesary even for lower dimensional systems, see for instance Ref. \cite{Entropy-new} and references therein.

Recently, the majorization relation was exploited to refine the uncertainty relation \cite{Majorization-1, Majorization-2}, of which the direct sum form usually has a better lower bound than the direct product one's \cite{Example-RPZ} and both of them still have rooms for further optimization \cite{Fei-improved, Maj-mix-sum}. A recent important progress in the study of uncertainty relation pertains to its generalization to incorporate the quantum entanglement \cite{Uncertainty-Memory-1}. The entanglement allows a violation of the uncertainty relation for particle $A$ if we have access to its entangled partner $B$. But how the multipartite entanglement effects the local uncertainty relation of each individual particles remains unclear, because the characterization of multipartite entanglement itself is already a complicated issue, in which the number of inequivalent classes turns out to be huge, whereas attainable via high order singular value decomposition (HOSVD) \cite{HOSVD-LU-1, HOSVD-LU-2, Wrapping-1, Multi-Class}.

In this paper we present a general scheme to incorporate the quantum uncertainty and multipartite entanglement: by exploring the statistical interpretation of quantum mechanics, we obtain majorization relations for the probability distribution and joint probability distribution of single particle and multipartite quantum states respectively. With this scheme, we solve the optimal lower bounds problem for entropic and majorization uncertainty relations of multiple observables, applicable to the general positive operator-valued measurements (POVM) and arbitrary mixed states. And the entropic uncertainty relations in the presence of multipartite entanglement are also obtained, where the lower bounds are found to be connected with the multipartite entanglement classes under local unitary equivalence. Hence, the characterization of uncertainty relation is now quantitatively related to the multipartite entanglement, which may enable the study of this fundamental relation in quantum mechanics to be extended to the more general situation, the many-body system.

\section{Majorization for single particle state}

In quantum theory, a physical observable is represented by a Hermitian operator. In the $N$-level discrete system, an observable $X$ is an $N\times N$ dimensional Hermitian matrix whose spectrum decomposition goes as
\begin{align}
X = u_x \Lambda_x u_x^{\dag} = \sum_{i=1}^N x_{i}|x_{i} \rangle \langle x_{i}| \; .
\end{align}
Here, $u_{x} = (|x_1\rangle, \cdots, |x_N\rangle)$ is unitary with $X|x_{i}\rangle=x_{i}|x_{i}\rangle$, and $\Lambda_x = \mathrm{diag}\{x_1, \cdots, x_N\}$ is the eigenvalue matrix. The quantum state $\rho$ of the system is also a Hermitian matrix
\begin{equation}
\rho = u \Lambda u^{\dag} = \sum_{i=1}^N \lambda_{i} |\phi_{i}\rangle \langle \phi_{i}| \; ,
\end{equation}
where $u = (|\phi_1\rangle, \cdots, |\phi_N\rangle)$ is a unitary matrix, and $\Lambda = \mathrm{diag}\{\lambda_1, \cdots,\lambda_N\}$ with $\lambda_1\geq \lambda_2 \geq \cdots \geq \lambda_N \geq 0$, the population rates at states $|\phi_i \rangle$,  and $\sum_{i=1}^N \lambda_{i}=1$. According to the statistical interpretation of quantum mechanics, when measuring $X$ over a quantum state $\rho$, we can only get its eigenvalue $x_{i}$ with a probability of $p_{i} = \langle x_{i}|\rho| x_{i}\rangle$. The probability distribution is then allowed to be expressed as a vector $\vec{p} = (p_1,\cdots, p_N)^{\mathrm{T}}$, where the superscript T means the matrix transposition. Then, we may define a set of Hermitian operators:
\begin{equation}
\mathcal{S}^{(x)}_n = \left\{ X_{n} \left| X_{n} = \sum_{\mu=1}^n |x_{i_{\mu}}\rangle \langle x_{i_{\mu}}|,\ \{i_{1},\cdots,i_{n} \}\subseteq \{1, \cdots, N\} \right.\right\}\ ,
\end{equation}
whose cardinality equals $\mathrm{C}_N^{n} = \frac{N!}{n!(N-n)!}$, because we may pick out $n$ different $|x_{i}\rangle$s from the total dimension of $N$($n\leq N$). From the definition, it is easy to find that the partial sum of the probability distribution $\vec{p}$ may write:
\begin{equation}
\sum_{i=1}^n p_{i} = \mathrm{Tr}[X_n \rho] \; . \label{partial-sums}
\end{equation}
Equation (\ref{partial-sums}) is also applicable to the general POVM, where the projection operators $|x_{i}\rangle \langle x_{i}|$ are replaced by positive semi-definite operators $M_{i}$ satisfying the normalization condition $\sum_{i} M_{i}^{\dag} M_{i} = \mathds{1}$ \cite{Quant-Infor-BOOK}.

The majorization between two tuples of real numbers, $\vec{a} \prec \vec{b}$ say, is defined as \cite{Majorization-Book}:
\begin{equation}
\sum_{i=1}^k {a}^{\downarrow}_i \leq \sum_{j=1}^k {b}^{\downarrow}_j \; , \; k\in \{1,\cdots, N\} \; ,
\end{equation}
where the superscript $\downarrow$ means that the components of vectors $\vec{a}$ and $\vec{b}$ are sorted in descending order, and the equality holds when $k=N$.

\begin{proposition}
In $N$-dimensional quantum system $\rho$, the measuring probabilities of arbitrary observables $X$, $Y$, and $Z$ satisfy the majorization relation:
\begin{equation}
\vec{p} \oplus \vec{q} \oplus \vec{r} \prec \vec{s}\; . \label{Single-Maj-two}
\end{equation}
Here, $p_{i} = \mathrm{Tr}[|x_i\rangle\langle x_i|\rho]$, $q_j= \mathrm{Tr}[ |y_j\rangle \langle y_j|\rho]$, and $r_k= \mathrm{Tr}[|z_k\rangle\langle z_k|\rho]$; $\vec{s} = (t_1,t_2-t_1, \cdots, t_{3N}- t_{3N-1})$ with $t_{\mu} = \max\,\{\vec{\lambda}^{\downarrow}_{\rho}\cdot \vec{\lambda}^{\downarrow}_{lmn} | l+m+n= \mu; l,m,n\in \{0,\cdots, N\}\}$. $\vec{\lambda}_{\rho}^{\downarrow} = (\lambda_{1},\cdots, \lambda_N)^{\mathrm{T}}$ is the eigenvalue vector of density matrix $\rho$ in descending order; $\vec{\lambda}_{lmn}^{\downarrow}$ is the eigenvalue vector of the Hermitian matrix sum $X_l +Y_m+Z_n$ in descending order, and $X_l, Y_m, Z_n \in \mathcal{S}_l^{(x)}, \mathcal{S}_m^{(y)},\mathcal{S}_n^{(z)}$ respectively. \label{Theorem-single-two}
\end{proposition}
\noindent {\bf Proof:} Since there exists obviously the relation
\begin{align}
\sum_{i,j,k=0}^{l,m,n} p_{i} + q_j + r_k  = \mathrm{Tr}[(X_{l}+Y_m+Z_n) \rho] \leq \vec{\lambda}^{\downarrow}_{\rho}\cdot \vec{\lambda}^{\downarrow}_{lmn}\ ,
\end{align}
where the equality is satisfied when eigenvalues of summed matrices $X_l +Y_m+Z_n$ are coincidentally in descending order with that of $\rho$, the majorization relation (\ref{Single-Maj-two}) then holds with $\mathcal{S}_0^{(x,y,z)} = \{0\}$. Q.E.D.

Note that from above Proof and equation (\ref{partial-sums}), it is obvious that the Proposition \ref{Theorem-single-two} is applicable to arbitrary number of operators and the general POVM measurements. And, it is worth emphasizing that (\ref{Single-Maj-two}) is an optimal relation, since there always exists the state $\rho$ for which the equality may be satisfied for given $l$, $m$, and $n$. Furthermore, it was noticed recently that the Horn's problem about the Hermitian matrix sum, specifically the majorization procedure, turns out to play a critical role in the study of the separability problem of the multipartite state \cite{Separability-1, Separability-2}.

A direct application of Proposition \ref{Theorem-single-two} leads to the following entropic uncertainty relation for multiple observables \cite{majorization}:
\begin{corollary}
For $M$ observables $X^{(i)}$, $i\in \{1, \cdots, M\}$, there exists the following entropic uncertainty relation
\begin{equation}
\sum_{i=1}^M H(X^{(i)}) \geq H(\vec{s}\,) \; . \label{equation-single-two-entropy}
\end{equation}
Here, the Shannon entropy $H(X^{(i)})$ is defined as $H(\vec{p}^{\,(i)})$ with $\vec{p}^{\,(i)}$ being the measuring probabilities of the $i$th observable $X^{(i)}$, and $\vec{s}$ is the same as in Proposition \ref{Theorem-single-two} satisfying $\bigoplus_{i=1}^M \vec{p}^{\,(i)} \prec \vec{s}$. \label{Corollary-single-two-entropy}
\end{corollary}
It should be remarked that though the vector $\vec{s}$ obtained from $t_{\mu}$ is optimized in the majorization uncertainty relation of Proposition \ref{Theorem-single-two}, the magnitude of $H(\vec{s}\,)$ may not be optimal in Corollary \ref{Corollary-single-two-entropy}, due to the reason that the maximum $t_{\mu}$ (here $\mu \in \{1,\cdots, MN\}$) is usually reached in different quantum states, as in vector $\vec{s}$. However, since the vector $\vec{s}$ and its component permutations comprise the outmost convex hull for the direct sum of vectors in equation (\ref{Single-Maj-two}), the following conjecture is reasonably establised:
\begin{conjecture}
The optimal lower bound of the entropic uncertainty relation in Proposition \ref{Corollary-single-two-entropy} may be obtained through one of the states maximizing the $t_{\mu}$, that is the $max\, \{\vec{\lambda}^{\downarrow}_{\rho} \cdot \vec{\lambda}^{\downarrow}_{i_1 \cdots i_M} | \sum_{j=1}^M i_j = \mu; i_j \in \{0, \cdots, N\}\}$, where $\mu\leq K$ with $t_{K-1} < M$ and $t_{K}=M$. \label{Conjecture-1-single}
\end{conjecture}
In exposition of above Proposition, Corollary, and Conjecture, following examples are helpful. Note, hereafter, the `log' implies the 2-based logarithm for the sake of comparison with results in the literature, and `$\ln$' stands for natural logarithm.

For illustration, consider two qubit observables $Z = u_z\Lambda_z u_z^{\dag}$ and $X_{\theta} = u_x\Lambda_x u_x^{\dag}$, with
\begin{equation}
u_z = (|z_1\rangle, |z_2\rangle) =
\begin{pmatrix}1 & 0 \\
0 & 1 \end{pmatrix} \; , \; u_x = (|x_1\rangle, |x_2\rangle) = \begin{pmatrix}
\cos\frac{\theta}{2} & \sin\frac{\theta}{2} \\
\sin\frac{\theta}{2} & -\cos\frac{\theta}{2}
\end{pmatrix} \; , \label{Example-z-x}
\end{equation}
where $\Lambda_{z} = \Lambda_x = \mathrm{diag}\{1,-1\}$, $0\leq \theta\leq \frac{\pi}{2}$. According to Proposition \ref{Theorem-single-two}, if $\vec{\lambda}_{\rho}^{\downarrow} = (\lambda_1,\lambda_2)^{\mathrm{T}}$ with $\lambda_1\geq \lambda_2\geq 0$ and $\lambda_1 + \lambda_2=1$, the magnitudes of $t_{\mu}$ are
\begin{align}
t_1 = \lambda_1 \; , \; t_2 = \lambda_1(1+ \cos \frac{\theta}{2}) + \lambda_2(1 - \cos \frac{\theta}{2}) \; , \; t_3 = 2\lambda_1 + \lambda_2 \; , \; t_4 = 2 \ .
\end{align}
Here, for example, $t_2$ may be obtained by $\mathrm{Tr}[(|z_1\rangle\langle z_1| + |x_1 \rangle \langle x_1|) \rho]$ with $\rho = \lambda_1 |t_2^{(+)}\rangle \langle t_2^{(+)}| + \lambda_2 |t_2^{(-)}\rangle \langle t_2^{(-)}|$ and $(|z_1\rangle \langle z_1| + |x_1\rangle \langle x_1|)|t_2^{(\pm)}\rangle = (1\pm \cos \frac{\theta}{2})|t_2^{(\pm)}\rangle$. The optimal vector $\vec{s}$ for the sum majorization uncertainty relation is then
\begin{equation}
\vec{p}_{z} \oplus \vec{q}_{x} \prec \vec{s} = (\lambda_1, \lambda_1\cos\frac{\theta}{2} + 2\lambda_2 \sin^2 \frac{\theta}{4}, 2\lambda_1 \sin^2\frac{\theta}{4} +\lambda_2\cos\frac{\theta}{2}, \lambda_2)^{\mathrm{T}} \; . \label{Example-mixed-s}
\end{equation}
Consider the case in pure state, $\lambda_1=1$ and $\lambda_2=0$, we have
\begin{equation}
\vec{p}_{z} \oplus \vec{q}_{x} \prec \vec{s} = (1,\cos\frac{\theta}{2}, 1- \cos\frac{\theta}{2}, 0)^{\mathrm{T}}\; . \label{pure-maj-s}
\end{equation}
Though $\vec{s}$ is optimal for equation (\ref{pure-maj-s}), $H(\vec{s}\,)$ is not the optimal lower bound for the sum $H(\vec{p}_z) + H(\vec{p}_x)$ according to Conjecture \ref{Conjecture-1-single}. The eigenvectors that maximize $t_{\mu}$ are
\begin{align}
t_1 & =1 \; \hspace{1.5cm}: \; \hspace{0.5cm}\{|z_1\rangle, |z_2\rangle, |x_1\rangle, |x_2\rangle\}\; , \\
t_2 & = 1 + \cos\frac{\theta}{2} \; : \; \hspace{0.5cm} \{\begin{pmatrix}\cos\frac{\theta}{4} \\ \sin\frac{\theta}{4}\end{pmatrix}, \begin{pmatrix}\sin\frac{\theta}{4} \\ -\cos\frac{\theta}{4}\end{pmatrix}\} \; , \\
t_3 & = 2 \;\hspace{1.5cm} : \; \hspace{0.5cm} \{ |z_1\rangle, |z_2\rangle, |x_1\rangle, |x_2\rangle \}\; ,
\end{align}
which give the probability distributions of $\vec{p}_z$ and $\vec{q}_x$
\begin{align}
t1\; , \; t_3 & \; : \; \vec{p}_z = (1,0)^{\mathrm{T}} \;, \; \vec{q}_{x} = (\cos^2\frac{\theta}{2}, \sin^2\frac{\theta}{2})^{\mathrm{T}} \; ,\\
t_2 &\; : \; \vec{p}_z = \vec{q}_{x} = (\cos^2\frac{\theta}{4}, \sin^2\frac{\theta}{4})^{\mathrm{T}} \; .
\end{align}
Hence, the corresponding Shannon entropies are
\begin{equation}
H(t_1) = H(t_3) = \displaystyle - \cos^2\frac{\theta}{2} \log(\cos^2\frac{\theta}{2}) - \sin^2\frac{\theta}{2} \log(\sin^2\frac{\theta}{2})
\end{equation}
and
\begin{equation}
H(t_2) = \displaystyle -2[\cos^2\frac{\theta}{4} \log(\cos^2\frac{\theta}{4}) + \sin^2\frac{\theta}{4} \log(\sin^2\frac{\theta}{4})] \ .
\end{equation}
Conjecture \ref{Conjecture-1-single} may be checked in typical $\theta$ values, like
\begin{align}
\theta = \frac{\pi}{2} & : \; H(\vec{s}\, ) \sim 0.872 \; , \; \boxed{H(t_{3})=1.000} \; , \; H(t_2) \sim 1.202 \; , \\
\theta = \frac{\pi}{3} & : \; H(\vec{s}\, ) \sim 0.568 \; , \; \boxed{H(t_2) \sim 0.709} \; , \; H(t_3) \sim 0.811 \\ \theta = \frac{\pi}{4} & :\;  H(\vec{s}\, ) \sim 0.388 \; , \; \boxed{H(t_{2}) \sim 0.467} \; , \; H(t_3) \sim 0.601\; , \\
\theta = \frac{\pi}{6} & : \; H(\vec{s}\, ) \sim 0.214 \; , \; \boxed{H(t_{2}) \sim 0.249} \; , \; H(t_3) \sim 0.355 \; .
\end{align}
The optimal lower bounds for entropic uncertainty relation (the boxed values) all agree with the analytical results for optimal entropic uncertainty relations of qubit systems \cite{Equiv-V-E}.

For two qutrit case, consider observables $X$ and $Y$ with orthonormal bases $X = \{|0\rangle, |1\rangle, |2\rangle\}$ and $Y = \{V|0\rangle, V|1\rangle, V|2\rangle\}$ \cite{Example-Coles}, where
\begin{align}
u_x & = (|x_1\rangle, |x_2\rangle, |x_3\rangle) = \begin{pmatrix}
1 & 0 & 0 \\
0 & 1 & 0 \\
0 & 0 & 1
\end{pmatrix} \;, \\
 u_y & = (|y_1\rangle, |y_2\rangle, |y_3\rangle) =
\begin{pmatrix}
\frac{1}{\sqrt{3}} & \frac{1}{\sqrt{3}} & \frac{1}{\sqrt{3}} \\
\frac{1}{\sqrt{2}} & 0 & -\frac{1}{\sqrt{2}} \\
\frac{1}{\sqrt{6}} & -\sqrt{\frac{2}{3}} & \frac{1}{\sqrt{6}}
\end{pmatrix} \; .
\end{align}
Similar to the qubit case we have $t_1=1$, $t_2 = 1+\frac{\sqrt{6}}{3}$, $t_3 = t_4=t_5=t_6=2$. Therefore the optimal bound for the sum majorization uncertainty relation is
\begin{equation}
\vec{p}_x \oplus \vec{p}_y \prec \vec{s}= (1,\frac{\sqrt{6}}{3}, 1-\frac{\sqrt{6}}{3},0,0,0)\; .
\end{equation}
The value $H(\vec{s}\,) \sim 0.688$ agrees with that of Ref. \cite{Example-RPZ} and is larger than $\sim 0.623$ of Ref. \cite{Example-Coles}. Similar to the qubit example, optimal lower bound for $H(X) + H(Y)$ is not $H(\vec{s}\,)$. According to Conjecture \ref{Conjecture-1-single}, by enumerating all the eigenvectors for $t_{\mu}$ we find that the optimal lower bound happens for the quantum state
\begin{align}
|\psi\rangle = (-\frac{1}{\sqrt{3}}, 0, \frac{\sqrt{2}}{\sqrt{3}})^{\mathrm{T}}\; ,
\end{align}
which gives $\vec{p}_x = (\frac{1}{3},0,\frac{2}{3})\; , \; \vec{q}_y = (0, 1, 0)$, and $H(X) +H(Y) \geq H(\vec{p}_x) +H(\vec{q}_y) \sim 0.918$.

\section{Uncertainty relation for multipartite states}

An $I_1\times \cdots \times I_M$ dimensional $M$-partite pure state $|\Psi\rangle = \sum_{i_1, \cdots, i_M=1}^{N} \psi_{i_1 \cdots i_M}| i_1, \cdots, i_M\rangle$ may be represented by a tensor $\Psi$ whose elements are $\psi_{i_1\cdots i_M}$. The HOSVD of $\Psi$ is given by \cite{HOSVD-LU-1, HOSVD-LU-2}
\begin{equation}
\Psi =  u^{(1)} \otimes   \cdots \otimes u^{(M)}\, \Omega\; .
\end{equation}
where $\Omega$ is an all orthogonal $M$-order tensor, named core tensor, characterizing the entanglement classes under local unitary equivalence. The $k$th mode unfolding of a high order tensor $\Psi$ is represented by $\Psi_{(k)}$, which is an $[I_k\times (I_{k+1}\cdots I_MI_1I_2\cdots I_{k-1})]$ dimensional matrix with elements $\psi_{i_k(i_{k+1}\cdots i_Mi_1i_2\cdots i_{k-1})}$. The one particle reduced density matrix is obtained by tracing out the rest particles
\begin{align}
\rho^{(k)} & = \mathrm{Tr}_{\neg k}[|\Psi\rangle \langle \Psi|] = \sum_{\substack{ i_1,\cdots,i_{k-1} \\ i_{k+1},\cdots, i_M }} \psi_{i_1 \cdots i_k \cdots i_M}^* \psi_{i_1 \cdots i'_k\cdots i_M}  |i'_k\rangle \langle i_k| \nonumber \\
& = u^{(k)} \Lambda^2_{\sigma^{(k)}} u^{(k)\dag} \; , \label{k-mod-singular-values}
\end{align}
where the sign $\neg k$ means the operation runs over all indices except the $k$th, and $ \Lambda_{\sigma^{(k)}} = \mathrm{diag}\{\sigma_{1}^{(k)},\cdots, \sigma_{I_k}^{(k)}\}$ with $\sigma_{i_k}^{(k)}$ being the $k$th mode singular values of the matrix unfolding $\Psi_{(k)}$.

The Hadamard product of two tensors is defined as $[A \circ B]_{i_1 \cdots i_M} =A_{i_1 \cdots i_M} B_{i_1 \cdots i_M}$. When measuring $X^{(k)}$ on particle $k$, the multivariate joint probability distribution of observing $x_{i_k}^{(k)}$ reads $P_{i_1 \cdots i_M} = P(x_{i_1}^{(1)}, \cdots, x_{i_M}^{(M)})$, which is also a $M$-order tensor with nonnegative elements, and may write like
\begin{equation}
P_{i_1\cdots i_M}
 = \left(\bigotimes_{k=1}^M (u_{x^{(k)}}^{\dag} u^{(k)})\, \Omega \right) \circ \left(\bigotimes_{k=1}^M (u_{x^{(k)}}^{\dag} u^{(k)})\, \Omega \right)^* \; . \label{Multi-Prob-Had}
\end{equation}
The above equation (\ref{Multi-Prob-Had}), characterized by the entanglement class of $\Omega$ under local unitary transformation, indicates that the probability distribution of observing $x_{i_k}^{(k)}$, $k\in \{1,\cdots,M\}$, equals to the modulus of $\psi'_{i_1\cdots i_M}\psi'^*_{i_1\cdots i_M}$, with $\Psi'$ being the quantum state $\Psi$ in the eigen basis of $X^{(k)}$, i.e. $\Psi'=\otimes_{k=1}^M u_{x^{(k)}}^{\dag} \Psi\ $, and the correlation and multipartite nonlocality are not independent natures of quantum state. The marginal probability distributions can then be read from the tensor $P$ as per summing over specified indices, i.e. $\vec{p}^{\,(k)} = \sum_{\neg k} P_{i_1\cdots i_k\cdots i_M}$ and $\vec{p}^{\,(\overline{ k})} = \sum_{i_k} P_{i_1\cdots i_k\cdots i_M}$, where the $k$th mode unfolding $P_{(k)}$ of $P$ may be regarded as a bipartite joint probability distribution.

For the joint probability distribution $P_{(k)}$, the Shannon mutual information and the classical correlation distance between the $k$th particle and the others are defined as
\begin{align}
\mathcal{I}(P_{(k)})& \equiv \mathcal{I}(k;\neg k) = \sum_{\mu,\nu} [P_{(k)}]_{\mu\nu} \log \frac{[P_{(k)}]_{\mu\nu}}{p^{(k)}_{\mu} p^{(\overline{k})}_{\nu}} \; , \\
\mathcal{C}(P_{(k)}) & \equiv ||[P_{(k)}]_{\mu\nu} - p^{(k)}_{\mu} p^{(\overline{k})}_{\nu}||_1 \; .
\end{align}
Here, the $L_{\alpha}$-norm between distributions is defined as $||p-q||_{\alpha} \equiv (\sum_j |p_j-q_j|^{\alpha})^{1/\alpha}$. Note, there exist a well-known relation $\mathcal{I}(P_{(k)}) \geq \frac{1}{2\ln 2} \mathcal{C}^2(P_{(k)})$ \cite{Cor-dist-Hall}.

\begin{lemma}
The marginal probability distribution $\vec{p}^{\,(k)}$ for measuring $X^{(k)}$ in the multipartite state $|\Psi\rangle$ is majorized by its $k$th mode singular values
\begin{equation}
\vec{p}^{\,(k)} \prec ( (\sigma_1^{(k)})^2, \cdots, (\sigma_{I_k}^{(k)})^2 )^{\mathrm{T}} \; . \label{maj-multiple-reduce}
\end{equation}
Of the joint probability distribution $P_{(k)}$, the mutual information between the $k$th particle and others satisfies
\begin{align}
\mathcal{I} \geq \frac{1}{2\ln 2} \mathcal{C}^2 \geq \frac{1}{2\ln 2}[||P_{(k)}||^2_2 - \sigma_1^2(P_{(k)})]\; , \label{Lemma-bound-mutual}
\end{align}
where $L_2$-norm reads $||P||_2^2 = \sum_{\mu,\nu} |P_{\mu\nu}|^2$, and $\sigma_1(\cdot)$ signifies the largest singular value. \label{lemma-2}
\end{lemma}

\noindent{\bf Proof:} Because the components of the probability vector $\vec{p}^{\, (k)}$ are $p^{(k)}_{i_k} = \langle x_{i_k}^{(k)}|\rho^{(k)} |x_{i_k}^{(k)}\rangle$, from (\ref{k-mod-singular-values}) we have
\begin{align}
\vec{p}^{\,(k)} & = \chi[u_x^{\dag} \rho^{(k)} u_x ] =  \chi[u_x^{\dagger}u^{(k)} \Lambda^2_{\sigma^{(k)}} u^{(k)\dagger} u_x ] \; ,
\end{align}
where $\chi[A] \equiv (A_{11},A_{22},\cdots, A_{NN})^{\mathrm{T}}$ denotes the operation that transforms the diagonal elements of a matrix into a vector. According to the Schur Theorem for Hermitian matrices, we then have  the majorized relation (\ref{maj-multiple-reduce}) (see Theorem 4.3.45 of Ref. \cite{Matrix-analysis}).

For classical correlation distance $\mathcal{C}(P)$, it obviously satisfies
\begin{align}
\mathcal{C}(P) \equiv \sum_{\mu,\nu=1}|P_{\mu\nu} - p_{\mu}q_{\nu}| \geq \sqrt{\sum_{\mu,\nu=1} |P_{\mu\nu}-p_{\mu}q_{\nu}|^2}  \; .
\end{align}
Here, $p_{\mu} = \sum_{\nu} P_{\mu\nu}$ and $q_{\mu} = \sum_{\mu} P_{\mu\nu}$ are marginal probability distributions. The quantity under the square root may be seen as the best rank-one approximation problem for matrix $P_{\mu\nu}$ (Theorem 7.4.1.3 in Ref. \cite{Matrix-analysis}), and hence
\begin{align}
\sum_{\mu,\nu=1} |P_{\mu\nu}-p_{\mu}q_{\nu}|^2 & \geq \sum_{i=1} \left[ \sigma_i(P)-\sigma(\vec{p}\vec{q}^{\,\mathrm{T}}) \right]^2 \geq \sum_{i=2} \sigma^2_i(P) \nonumber \\
& = ||P||^2_{2}- \sigma^2_{1}(P) \; , \label{norm-singular}
\end{align}
where $||P||_2=||P||_{\mathrm{F}}$ is also named as Frobenius norm. Q.E.D.

The Lemma \ref{lemma-2} tells that the local measurement on observable $X^{(k)}$ can only have more spread distribution than the singular value squared, the correlation of the $k$th particle with others is determined by singular values of the joint probability distribution $P_{(k)}$, which is different from but related to the singular values of core tensor $\Omega$ of the quantum state $\Psi$. A direct application of Lemma \ref{lemma-2} yields the following uncertainty relation in the presence of bipartite entanglement.
\begin{proposition}
For joint measurements of $X=u_x\Lambda_x u_x^{\dag}$ and $Z=u_z\Lambda_z u_z^{\dag}$ on particle $A$, and $Y=u_y \Lambda_y u_y^{\dag}$ and $W=u_w\Lambda_w u_w^{\dag}$ on particle $B$, the correlation between the particles $A$ and $B$ reduces the uncertainties by an amount of $\mathcal{I}_{AB}$, that is
\begin{align}
H(X|Z) + H(Y|W) &= H(X) + H(Y) - \mathcal{I}_{AB} \; , \label{Theorem-2-equ-reduce}
\end{align}
where $\mathcal{I}_{AB} = \mathcal{I}(P) + \mathcal{I}(Q)\geq \frac{1}{2\ln 2}\left[||P||_{2}^2 -\sigma_1^2(P) + ||Q||_{2}^2 - \sigma_1^2(Q)\right] \geq 0$,
with $P = (u_x^{\dag}\otimes u_z^{\dag} \Psi) \circ (u_x^{\dag}\otimes u_z^{\dag} \Psi)^*$, $Q = (u_y^{\dag}\otimes u_w^{\dag} \Psi) \circ (u_y^{\dag}\otimes u_w^{\dag} \Psi)^*$. $\mathcal{I}_{AB}$ is zero if and only if the joint probability distributions $P$ and $Q$ both are direct products of two marginal distributions. \label{Theorem-2}
\end{proposition}
This Proposition follows from the relation between the Shannon conditional entropy and mutual information $H(X|Z)=H(X)- \mathcal{I}(X;Z)$. We shall show that Proposition \ref{Theorem-2} leads to a universal entropic uncertainty relation in the presence of quantum memory, which is quite easy for experimental verification.

Existing result for the entropic uncertainty relation in the presence of quantum memory is \cite{Uncertainty-Memory-1}
\begin{equation}
S(X|B) + S(Y|B) \geq c_{\mathrm{s}} + S(A|B) \; . \label{entropy-memory}
\end{equation}
Here the Von Neumann entropy $S(\rho) \equiv -\mathrm{Tr}[\rho \log \rho]$, $S(X|B) = S(\rho_{XB}) - S(\rho_B)$, with $\rho_{XB} = (\mathcal{X} \otimes \mathds{1}) \rho_{AB}$ and $\mathcal{X}(\cdot) = \sum_j |x_j\rangle \langle x_j|(\cdot)|x_j\rangle \langle x_j|$. Though a large number of different lower bounds exist for the entropic uncertainty relation, only a few of them can be taken into equation (\ref{entropy-memory}) as $c_{\mathrm{s}}$ \cite{Fei-improved, UR-M-M}. This prevents the quantity $S(A|B)$ from being regarded as a reduction of the uncertainty for the variant lower bounded entropic uncertainty relations. Contrarily, equation (\ref{Theorem-2-equ-reduce}) predicts that
\begin{equation}
H(X|Z) + H(Y|W) \geq  c_{\mathrm{h}} - \mathcal{I}_{AB} \; , \label{entropy-memory-Shannon}
\end{equation}
where all the lower bounds for the entropic uncertainty relation in the literature can be taken into equation (\ref{entropy-memory-Shannon}) as $c_{\mathrm{h}}$ directly. In this sense, $\mathcal{I}_{AB}$ represents a universal measure of the reduction of the uncertainty for the incompatible observables. To be specific, for the state $|\Psi_{AB}\rangle = \frac{1}{\sqrt{2}}(|00\rangle + |11\rangle)$ with observables $Z$ and $X_{\frac{\pi}{4}}$ ( see equation (\ref{Example-z-x})) on $A$ and $Z'$ and $X'_{\frac{\pi}{4}}$ on particle $B$, the lower bounds predict by equations (\ref{entropy-memory}) and (\ref{entropy-memory-Shannon}) are
\begin{eqnarray}
c_{\mathrm{s}}+ S(A|B) & = & c_{\mathrm{s}} - 1  = \log(\frac{1}{\cos^2\frac{\pi}{8}}) -1 \sim -0.772 \; , \label{Example-qubit-nat} \\
c_{\mathrm{h}} - \mathcal{I}_{AB} & = & H_{\mathrm{bin}}(\cos^2\frac{\alpha}{2}) + H_{\mathrm{bin}}(\cos^2 \frac{\beta}{2}) \geq 0 \; . \label{Example-univ-lower}
\end{eqnarray}
Here, $c_{\mathrm{s}}$ follows from Ref. \cite{Example-Coles}, $\alpha$($\beta$) is the angle between the measurement basis of $Z$($Z'$) and $X_{\frac{\pi}{4}}$($X_{\frac{\pi}{4}}'$), and the binary Shannon entropy $H_{\mathrm{bin}}(p) \equiv -p\log p -(1-p)\log(1-p)$. The lower bound in equation (\ref{Example-qubit-nat}) is a minus value, thus cannot provide us much insights into the uncertainty of incompatible observables on particle $A$. While according to equation (\ref{Example-univ-lower}), the uncertainty of the incompatible observables $Z$ and $X_{\frac{\pi}{4}}$ could be reduced to zero, by appropriately chosen $Z'$ and $X'_{\frac{\pi}{4}}$ on particle $B$, i.e. $\alpha, \beta = 0,\pi$ which corresponds to $Z' =\pm Z$ and $X_{\frac{\pi}{4}}' = \pm X_{\frac{\pi}{4}}$.

Another advantage of uncertainty relation (\ref{entropy-memory-Shannon}) is its experimental verification. If we only know the eigenvalues of the reduced density matrix of $\rho_A$, i.e. $\lambda_1 \geq \lambda_2>0$ with $\lambda_1+\lambda_2=1$, then lower bound $H(\vec{s}\,)$ of $H(Z) + H(X_{\theta})$ may be obtained according to equation (\ref{Example-mixed-s}), which is the best analytical result for mixed states as far as we know. For example, when $\lambda_1=3/4$, $\lambda_2=1/4$, and $\theta=\pi/3$, we have $H(\vec{s}\,) \sim 1.712$, larger than that of $B_{\mathrm{MU}}+H(A) \sim 1.226$ in Ref. \cite{UR-M-M}. With this experimental configuration, the uncertainty relation (\ref{entropy-memory-Shannon})  reads
\begin{equation}
H(Z|Z') + H(X_{\frac{\pi}{3}}|X'_{\frac{\pi}{3}}) \geq 1.712 -\mathcal{I}_{AB} \; . \label{Experimental-Sugg}
\end{equation}
Here the three quantities $H(Z|Z')$, $H(X_{\frac{\pi}{3}}|X'_{\frac{\pi}{3}})$, and $\mathcal{I}_{AB}$ involve only the coincidence counting rates, unlike that of equation (\ref{entropy-memory}) whose verification usually needs additional quantum state tomography \cite{UR-M-Exp-1,UR-M-Exp-2}.

\begin{corollary}
The uncertainty in measuring $X$ on particle $A$, who correlates with particles $B$ and $C$, has the following relation
\begin{equation}
H(X|Y,Z) = H(X) + \mathcal{I}(Y;Z) - \mathcal{T}(X;Y;Z) \; ,
\end{equation}
where $Y$ and $Z$ are the measurements performed on particle $B$ and $C$; $\mathcal{T}(X;Y;Z) \equiv H[P(x,y,z)||p(x)p(y)p(y)]$ is the total correlation equal to the relative entropy of joint probability distribution $P(x,y,z)$ to the product of its marginal ones. \label{Corollary-triple-entropy}
\end{corollary}
The Corollary \ref{Corollary-triple-entropy} is quite apparent from classical information theory, and it reveals that knowing only pairwise information is usually inadequate in reducing the local uncertainty of a multipartite state. The evaluation of the total correlation $\mathcal{T}(X;Y;Z)$ involves the estimation of a high order tensor with rank one approximations \cite{HOSVD-siam}, and is closely related to the entanglement classes under local unitary equivalence \cite{HOSVD-LU-1, HOSVD-LU-2}.

\section{Conclusions}

We develop in this paper an explicit entropy relation for quantum uncertainty and entanglement, the two extraordinary characters of quantum theory, for arbitrary multipartite system. By exploiting the majorization relation for diagonal entries and eigenvalues of the Hermitian matrix, we solve the optimization problems in state-independent majorization and entropic uncertainty relation. The uncertainty relation in the presence of multipartite entanglement is also constructed within the scheme, where the lower bound for joint probability distribution of incompatible obervables is found to be determined by the measurement basis and entanglement class under the local unitary equivalence. The result establishes a direct relationship between quantum uncertainty and multipartite entanglement, and is instructive to the further understanding of other nonlocal phenomena in mixed system, like separability of quantum states \cite{Separability-1,Separability-2} and quantum steering \cite{Nonlocal-via-UR}.

\section*{Acknowledgements}

\noindent
This work was supported in part by the Ministry of Science and Technology of the Peoples' Republic of China(2015CB856703); by the Strategic Priority Research Program of the Chinese Academy of Sciences, Grant No.XDB23030100; and by the National Natural Science Foundation of China(NSFC) under the grant 11635009.

\end{document}